\documentclass[preprint]{aastex7}
\shorttitle{Quadruple SN 2025wny: Lessons for LSST Alerts}
\shortauthors{Schechter, Lu and Hernandez}

\begin{document}

\title{The Quadruply Lensed Supernova SN 2025wny: Implications for LSST}

\author[0000-0002-5665-4172]{Paul L. Schechter}
\email{schech@mit.edu}
\affiliation{MIT Kavli Institute and Department of Physics, 77 Massachusetts Ave, Cambridge, MA, 02139,USA}

\author[0009-0001-2535-5735]{Michael Zeng Lu}
\email{mzlu@mit.edu}
\affiliation{MIT Kavli Institute and Department of Physics, 77 Massachusetts Ave, Cambridge, MA, 02139,USA}

\author[0009-0005-1678-8722]{Christopher Hern\'andez}
\email{CHH226@pitt.edu}
\affiliation{Department of Physics and Astronomy, University of Pittsburgh, 3941 O’Hara St, Pittsburgh 15260, USA}

\correspondingauthor{schech@mit.edu}

\begin{abstract}

Lensed supernovae (SNae) are among the most eagerly anticipated
transients expected from the Legacy Survey of Space and Time (LSST).
Quadruply lensed SNae permit more highly constrained models than
``mere" doubles.  The quadruply lensed SN 2025wny offers multiple
lessons on how one might respond to an alert. The full benefits of
such rare events are best achieved with {\it immediate} spectroscopic
and photometric followup, within hours rather than days.  This in turn
requires on-the-fly modeling to predict the position(s) and magnitudes
of trailing images and to ``pre-cover" any leading images that might
have been too faint to trigger an alert and that cannot be detected in
the triggering exposure.  This paper sets out a proposed protocol for
exploiting similar alerts.  A list of quadruply lensed candidate hosts
must first be supplied in advance to one or more brokers, along with on-the-fly
software (an example of which is given) to determine whether an SN
near an incipient host is strongly lensed, and whether quadruply or
doubly.  The brokers would then broadcast the positions and time
delays (or ``pre-lays") that permit ``pre-covery'' of leading images,
``re-covery'' of trailing images, and possibly, extraction of a rough lightcurve
from prior LSST exposures.  The scheme is illustrated (and some
potential problems identified) using preliminary data for SN 2025wny
presented by three independent teams.  It employs software based on the
geometric Witt-Wynne lens model and Falor's exact, forward,
differentiable solution thereof.
  
\end{abstract}




  
\section{Introduction}

The quadruply lensed supernova SN 2025wny was auspiciously discovered
(S.~Taubenberger et al. 2025, J.~Johansson et al. 2025) just months before the official start
of the Legacy Survey of Space and Time (LSST) of the Vera C.~Rubin
Observatory.   Systems of this sort are among the most keenly anticipated
LSST transients.

Five weeks passed between the identification of SN 2025wyn as a
candidate transient in the course of a routine survey with the Zwicky
Transient Facility \citep{Bellm} on 2025 August 29 and the report of
four distinct images by \citet{Wise} on October 3.  Only with the
detection on September 20 of two additional images beyond the original
triggering image did the pace of follow-up accelerate.  

A fourth image was discovered on September 29.  A Target of
Opportunity (TOO) program with HST carried out on 2025 October 20 and
November 30.  Another TOO program, with JWST, was
carried out on 2025 November 19 and 22.

A timeline of the observations and postings through October 27,
extracted from arXiv submissions that day by \citet{Taubenberger} and
J. Johannson et al. (2025) is presented in Appendix~\ref{app:timeline}.
Some of the many circumstances that might delay the confirmation of a system
as a quadruply lensed supernova may be gleaned from that timeline.

In \S 2 below, with the benefit of hindsight, we propose a generic
protocol for proceeding from the issuing of a ``time domain event alert packet''
to the reconstruction of the system model using the observations
of the supernova images.

The issuance of an alert packet is just the first of several
successive stages in this process.

In \S 3 we use SN 2025wny to illustrate how our proposed protocol
might have been applied.  We relegate discussion of the particular
parameters and predictions that a broker might have broadcast for this
alert to an appendix.  The word ``broker'' refers to a software system designed to ingest, process, and distribute astronomical alerts from the LSST and other surveys to the public.

The central and least familiar component of our proposed protocol is
the on-the-fly propagation by the broker of the alert coordinates
through a pre-computed lens model derived from an incipient host
(known in advance to be quadruply lensed) to either corroborate or
rebut the predicted coordinates of additional images. This broker task
is considerably more complex than establishing proximity to a target,
and perhaps more complex than linking alerts for Solar System Objects.

The observed position of the alert must first be traced through the
pre-computed model back to the redshift of the host and then projected
forward to give positions for all of the images expected from the
event.  For example, the host of SN 2025wny, PS1J0716+3821, was known
to be quadruple \citep{Canameras}.  Even so, the reported transient
might have been sufficiently offset from the center of the host to
produce just two images, or perhaps only one.  The predictions from
this propagation -- the number of images, their positions,
magnifications and time delays -- serve as the basis for
an efficient and effective follow-up campaign.

Throughout this work we employ the Witt-Wynne geometric prediction
\citep{Wynne} for the positions of the images produced by a
singular isothermal elliptical potential. The suitability of this model
to on-the-fly computation  compensates, for the present purposes,
for its mild shortcomings.

\vfill
\eject

\section{Generic Steps Toward the Re-covery of Quadruply Lensed Supernovae}

\subsection{Staged response to alerts near known lensed galaxies}

Here we enumerate the stages of our proposed protocol, with the caveat
that some stages may proceed in parallel.  Positive numbers
are used for those stages that follow the alert.\footnote{Our terminology differs in several ways from that adopted for the LSST
Data Management System's Alert Production pipeline.  We use
``triggered'' rather than ``detected'' to refer to sources with a
signal to noise ratio, SNR $>$ 5 in positive or negative flux by the
Difference Image Analysis (DIA) pipeline. We characterize an alert
that meets the conditions provided by a user as ``flagged'' instead of
having ``passed'' through its ``filter''.  We refer to the undeflected
(and unseen) supernova as the ``source'', and the observed
manifestations of the supernova ``images''.  By contrast LSST refers
to the latter as ``sources'' and refers to processed exposures as
images.}

\bigskip
{\baselineskip 0pt
\begin{itemize}
\item[\bf-1.] ``modeling'' of incipient host, known to be quaruply lensed
\item[\bf0.] DIA alert
\item[\bf1.]``flagging'' of alert, based on proximity
\item[\bf2.]``propagation'' of alert coordinates through host model
\item[\bf3.]``pre-/re-covery'' search for additional images using alert packet and/or ancillary observations
\item[\bf4.]``corroboration'' of predicted image positions
\item[\bf5.]``phasing'' of corroborated images using alert packet, ancillary observations or subseqent LSST imaging
\item[\bf6.]``verification'' as a supernova using ancillary spectroscopy
\item[\bf7.]``reconstruction'' incorporating SN images into an improved model
\end{itemize}
}
\bigskip

Following the triggering of an LSST alert, the first step is its
association, by virtue of angular proximity, with an entry on a
pre-selected list of objects of interest provided to an LSST community
alert broker.  In the present case these would be incipient host
galaxies that appear to be quadruply lensed.\footnote{ The list might
also include doubly lensed incipient hosts.  As a rule these will have
smaller (and more uncertain) diamond caustics  than their
quadruple cousins but an offset SN might nevertheless fall within one.
} We designate such an associated system as ``flagged."  The broker
might, at this point, broadcast the flagging of an alert.  But ideally
the will broker carry out the ``propagation'' step as well, for which
purpose the broker must be supplied with the gravitational potential
of the lens, pre-computed from the four host images.

``Propagation" of the alert position {\it back} to the redshift of the
incipient host requires the pre-computed lens potential.  In general,
the supernova will not be coincident with the host's center.  One must
therefore propagate the offset SN position {\it forward} through the
model to the plane of the lensing galaxy, predicting the expected 
number of additional images (three, one or none),
and their positions, magnifications, and time
delays.  The broker then transmits these predictions to interested parties for
immediate follow-up.

Next, the broker would check to see if an additional trigger lies
close to one of the three positions predicted for the other images of
the supernova.  Our hope would be that our list of incipient quads
would qualify for the limited list of targets that generate alerts at
a lower threshold, say {\it trans SNR = 3} rather than the global
threshold.  If a second triggering event is detected where predicted,
the system qualifies as ``corroborated,'' the broker  publishes
the confirming positions and magnitudes as well.  The broker also carries out
forced photometry on cutouts from previous nights,``pre-covery,''
which might also corroborate the predictions.  If corroboration is accomplished
using the alert packet, follow-up observations would in principle
be possible within hours of the alert (perhaps even minutes)
and would not be delayed until after the 80 hour ``embargo.''

Rebuttals must be carefully qualified since the SN images are subject
to micro-lensing, as may have been the case with SN2025mkn
\citep{lemon}.  Micro-lensing is less strong for positive parity images,
in particular the leading image, \citep{Weisenbach}, so the rebuttals
may gain force when the shape of the lightcurve is better known.

We take 3 epochs to be the minimum needed for phasing, of which
one must be before maximum and another must be after.  This puts a
premium on obtaining an {\it immdediate} a high cadence light curve on
the triggering transient, to take full advantage of the sparser
lightcurve of a leading image, and of the lower S/N in the ``re-covery''
of a trailing image.  Verification spectroscopy is also best carried
out {\it immediately} while the triggering image is still bright.

There are circumstances, involving either micro-lensing of the SN
images (as discussed above) or discoveries made near sunset and
sunrise, under which only two or three of the images of a quad will be
detected.  While such incompletely pre-/re-covered systems are still
of some use, completely pre-/re-covered systems permit stronger
science conclusions.

The image positions, time delays and magnifications predicted using
the gravitational potential of the lens are by nature uncertain,
reflecting the approximate nature of models and the signal-to-noise of
the data that constrain them.  The model can be reconstructed as each
successive SN image is recovered, and again when the supernova has faded,
permitting deeper and higher resolution imaging of the host galaxy.
A partial reconstruction might improve a pre-covery or re-covery,
by improving the constraints on the predicted position of a missing
image.

\section{Quad SN recovery protocol applied to SN 2025wny}

\begin{figure*}[ht!]
  \plotone{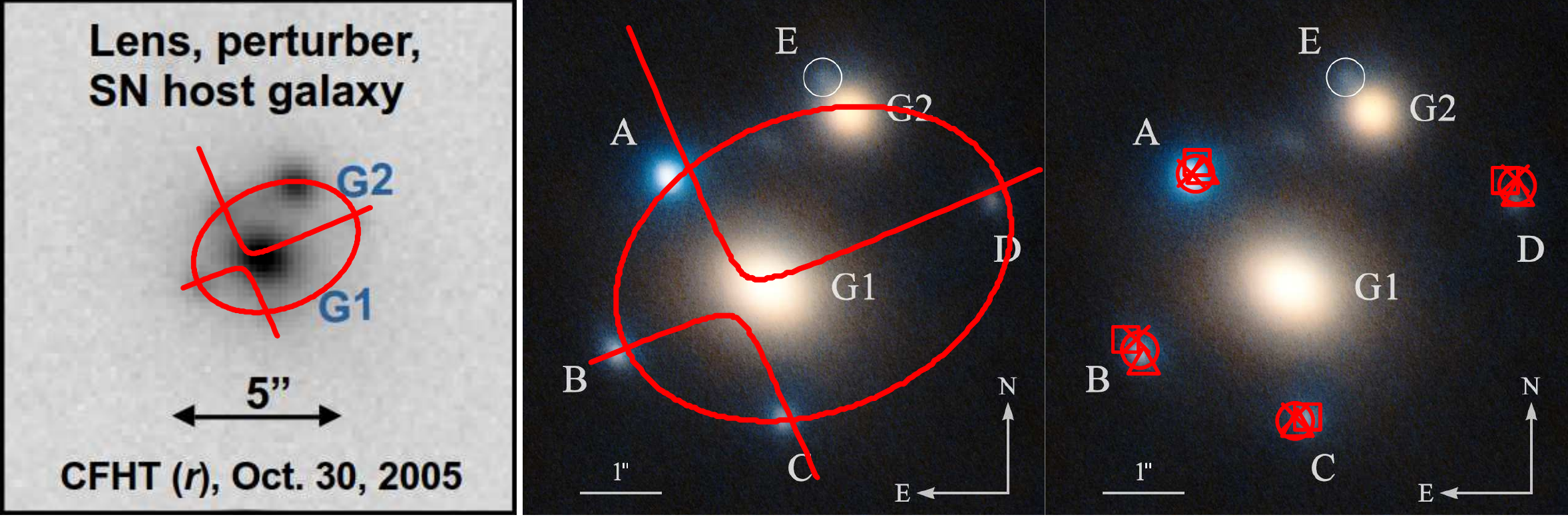}
  \caption{The left panel shows our geometric lens model overlayed on a pre-tranisent exposure of the quadruply lensed host galaxy, taken with the CFHT.  The western-most image of the $z = 2$ host is less highly magnified than the other three.  The model fit to the  four host positions puts them where Witt's hyperbola intersects     Wynne's ellipse.  The central panel shows the geometric lens model overlayed on an LBT $JK$  composite of SN 2025wny as observed on  2025
    November 25-26.  The right panel shows the positions predicted for the three {\it other} images obtained treating $A, B, C$ and $D$ (represented by an ``X'', an open
    circle, and open square and a triangle, respectively) as the triggering image.  The cutouts are reproduced from Figures 3 of \citet{Taubenberger} and 1 of \citet{Ecker}. }
\label{fig:hostSNpred}    
\end{figure*}

\subsection{A counterfactual triggering exposure}

Many of the virtues and shortcomings of the protocol proposed  in Section 2
are illustrated by applying it to the case of SN 2025wny.  To
simplify our presentation, we allow ourselves some historical latitude and treat
adaptive optics images obtained with the Large Binocular Telescope (LBT) 
on November 25-26 as the trigger for the alert,
rather than the ZTF direct exposure of August 29 that actually
triggered it (as tabulated in Appendix \ref{app:timeline}).  We discuss the
consequences of this choice in \S \ref{subsec:ZTF}.

The highest hurdle in applying our protocol is that numerical values for
the positions of the incipient host images are absent from both of
the discovery papers.  We circumvented this by measuring the positions
with a cursor on the figures supplied by the authors.  The uncertainties
in these  cursor measurements suffice to produce a reasonably good Witt-Wynne
model, as seen from its overlay on the CFHT cutout.  The positions
of $G1$ and $G2$ with respect to the model were likewise obtained with our
cursor.

For an early draft of this paper, we similarly measured the positions
of the four supernova images using the COLIBRI direct image in Figure
3 of \citet{Taubenberger}.  With the subsequent posting of
\citep{Ecker}, we the used the LBT $K$ band tabulated in their Table
1.  The differences between using those early measurements and LBT
measurements used in constructing the rightmost panel in Figure
\ref{fig:hostSNpred} were small.  The rightmost panel shows the predicted
positions obtained by treating each of the LBT positions, in turn, as
the triggering images. The scatter is roughly $0\farcs1$.

As the present paper was nearing completion, adaptive optics exposures
of SN 2025wny carried out with the Keck NIRC2 were reported by
\citet{Storfer}.  These were in substantial agreement, $0\farcs01$
rms, with the LBT observations apart from small differences in the
scaling and rotation of the field.

A common feature of all four arXiv postings is that {\it none} of them
report predicted time delays.  \citet{Storfer} come closest saying
``While we do not report time-delays in this work, it is worth noting
that images A and C (and E) are roughly in-phase with relative
time-delays between them on the order of a few days.''  This is
qualitatively consistent with the seven-day difference reported in
Table \ref{tab:modSTUFF}.  In the abstract of the same paper they say
SN 2025wny is``...  the first lensed supernova in a galaxy-scale
system that is suitable for time-delay cosmography studies.''

Is the time delay of image D {\it not} worth noting?  Unless we are
badly mistaken, it leads A and C by several months, which could well
render the system unsuitable for time-delay cosmography.

\citet{Storfer} say the ``The strategies developed to characterize
galaxy-scale lensed SNe like SN Winny are being deployed for the first
time. As new lensed SNe are discovered, this process will continue to
be refined.''  Our proposed protocol is one such possible refinement.

\subsection{Shortcomings of the singular isothermal elliptical potential}

Overplotted on the leftmost and central panels of Figure
\ref{fig:hostSNpred} is a geometric representation of a singular
isothermal elliptical model for the lens potential \citep{Wynne}.  The
model puts the four images where Wynne's ellipse intersects the
rectangular hyperbola derived by \citet{Witt}.

The model host positions are reported in Table \ref{tab:modSTUFF}.
Deviations of the model host positions from those observed are to be
expected, given the presence of two lensing galaxies rather than one.
For the purpose of comparison with future models (to be computed from
observations after the SN has faded) we give the parameters of our
model in Appendix \ref{app:WittWynne}.

\begin{deluxetable*}{lrrrr}\label{tab:modSTUFF}
\tablecolumns{5} \tablewidth{720pt} \tablecaption{Model Derived
  Quantities} \tablehead{ \colhead{} & \colhead{$\Delta x(\arcsec)$} &
  \colhead{$\Delta y(\arcsec)$} & \colhead{$\Delta t(days)$} &
  \colhead{$\mu$} \\ } \startdata Host Image A & -1.05 & +1.29 & 168 &
-1.15 \\ Host Image B & -1.81 & -0.87 & 132 & +2.70 \\ Host Image C &
+0.18 & -1.73 & 161 & -1.29 \\ Host Image D & +2.73 & +1.09 & 0 &
+1.75 \\ Lens G1 & +0.04 & -0.10 & & \\ Lens G2 & +1.06 & +2.08 & &
\\ \enddata \tablenotetext{$*$}{All positions are relative to the
  derived center of the model, positve to W and N.  Host positions are
  predicted. Lens positions are measured.}
\end{deluxetable*}

\citet{LuhtaruII} have noted that for most quadruple quasars lensed by
relatively isolated galaxies, the longer branch of Witt's hyperbola, as
predicted, passed quite close to the position of the lensing galaxy.
But the longer branch in Figure \ref{fig:hostSNpred} is substantially
offset from the center of G1, reflecting the influence of G2.
Moreover, \citet{Ecker} identified a fifth image in the LBT data,
$E$, slightly offset from $G2$, that the Witt-Wynne model does not
produce.

One must weigh these model shortcomings  against its success in
using one supernova position to predict (or postdict) the other three.
The central panel of Figure \ref{fig:hostSNpred} shows the same
geometric model overlayed on the LBT image.  Though none of the four
supernova images lie at the intersections of  Witt's hyperbola and
Wynne's ellipse,  their offsets are nearly identical to the host
offsets  in the CFHT image.  It would appear that the
supernova was close to the center of the host.

A hypothetical broker would broadcast that an alert had been
``flagged," because a) the transient was close to a known incipient host
and b) the transient had been propagated back to the redshift of the host
and then forward to the redshift of the lens, predicting (or in this
case postdicting) the positions of three additional images close to
the corresponding images of the host.  Spectroscopic ``verification"
would be undertaken by the broker's clients.

The LBT image successfully corroborates the model predictions.  We
note that the \citep{Taubenberger} earlier COLIBRI image likewise
corroborated the model.  The September 20 Liverpool exposure reported
by \citep{Johansson} might have corroborated our model but our
attention had not yet been drawn to SN2025wny.  With the two arXiv
postings on October 27 we use the best available from each to produce
to model the host and measure the SN image positions.  The ZTF
discovery and pre-discovery exposures might also have corroborated our
model, since the threshhold demanded for corroboration need not be as
high as the threshhold for discovery.

\subsection{Sufficiency of the Witt-Wynne model}

In the rightmost panel of Figure \ref{fig:hostSNpred} four points are
clustered around each of the four SN images.  One coincides
precisely with that image and the other three are located at the
positions for the SN predicted (or postdicted) from the other three
images.  The scatter is small and ought to suffice for the
``re-covery'' and ``pre-covery'' of the other images, the
above-mentioned imperfections in the Witt-Wynne model notwithstanding.

\subsection{ZTF25abnjznp:  the actual trigger}\label{subsec:ZTF}

For simplicity of presentation, we have chosen to treat the LBT composite
of 2025 November 25-26 as the trigger of our alert rather than the ZTF exposure
of August 29 that in fact led to the SN 2025wny alert (see Appendix
\ref{app:timeline}).  Under our proposed protocol  ZTF25abnjznp would have
immediately been ``flagged" as associated with the quadruply lensed
host, PS1J0716+3821.

A pre-computed lens model derived from the host would then have been
applied to the measured position of ZTF25abnjznp, and predictions made
for the other images.  If the error in the ZTF position were
sufficiently small, the model would have predicted four images and the
broker would have reported the system as a ``propagated" quad.\footnote{More
generally, one might sample the ZTF positional error ellipse and
compute a probability that the transient lies within the diamond
caustic, and similarly for the radial caustic.  But for the present we
assume that if the signal-to-noise is high enough to generate an
alert, it will also be high enough to produce a relatively accurate
position.}

As the ZTF pixels are 4 times bigger than the LSST pixels, one
might not expect the four SN images to be resolved in a ZTF image.
But one could nonetheless have used the position of the ZTF transient
to generate positions for additional images using the pre-computed
lens model.  Forced photometry might then have been carried out at
those positions, testing for significant excess flux.  Forced photometry
is part of the DIA toolkit, and is an included field in the LSST alert
packet schema.

The COLIBRI and Maidanak exposures shown by \citet{Taubenberger}
suggest that image A was 1-2 magnitudes brighter than images B and C.
By contrast, in our geometric model image B is twice as highly
magnified as either A or B (as shown in Appendix \ref{app:WittWynne}).
This might be due either to micro-lensing or to the double nature of
the lens galaxy system.  In either case, forced photometry might not
have yielded a significant detection of images B-D on the ZTF
discovery image.  It might, however, have yielded a significant
detection on the September 2nd Liverpool Telescope image described
in \citet{Johansson}.

\section {Beyond SN 2025wny}\label{sec:pg1115}

The discovery of quadruply lensed supernovae with LSST will be subject to
a great many selection effects, some of them reflecting details of
the alert system and others reflecting the discovery process and
modeling of the lensed incipient hosts.  One might reasonably
expect many of the incipient systems to have image configurations
much like that of the first quadruply lensed quasar, PG1115+080
\citep{Weymann}, which has a close pair of highly magnified images.
If most LSST alerts are close to the edge of detection, systems
with highly magnified images will preferentially generate alerts.

It is not an accident that the highly magnified images in PG1115+080
are separated by only $0\farcs48$ \citep{Vanderriest} -- it is characteristic of pairs of
images caused by so-called folds in the lensing caustic to be a) magnified
by the same factor, which b) is inversely proportional to their separation
\citep{Blandford}.  The time delay between the images is proportional to the square
of their separation, and in the case of PG 1115+080 is several hours.

If a supernova exploded in an incipient host with the same
configuration as PG 1115+080, it might produce a pair of blended
triggering images whose centroid would lie somewhere on an arc passing
through the two host images.  The moments and lengths calculated
by the DIA and included in the alert package may suffice to compute
the relative fluxes of the close pair and the source position.  The
DIA may succeed in splitting triggering image.  And if the incipient hosts
are included in limited sample of  for which reduced threshold alerts  {\it trans SNR = 3}
are distributed, there may be a second alert from the two less highly magnified
images.  In the worst case,  one would need to wait out
the 80 hour embargo to obtain forced photometry on the prompt pipeline
image, delaying the start of follow-up observations.

We anticipate refining of our protocol to deal with cases like PG 1115+080
following its initial implementation.  Among the refinements will be the
inclusion of a StarHASH designation \citep{Killestein} for each flagged system.

\section{Summary}

We have proposed a protocol to apply to LSST-like alerts 
to accelerate the identification and recovery of quadruply lensed
supernovae, and have demonstrated how it might have worked in
the case of SN 2025wny.  The
linchpin of the protocol is a pre-computed model for the lensing
gravitational potential obtained from a quadruply lensed incipient
host.  That model is then applied, on-the-fly, to propagate the observed
position of the transient backward to the source and then forward 
to compute the number of additional images of the transient and their
expected positions, magnifications and time delays.

The on-the-fly calculations can be carried out quickly and straightforwardly
using the Witt-Wynne geometric model for the singular isothermal elliptical
potential \citep{Wynne}, for which \citet{Falor} present an exact forward
solution for the positions of the images of a supernova at the redshift of
the host.

We distinguish seven stages on the way to confirming and recovering a
quadruply lensed supernova: flagging (as close
to an incipient host), propagation (projecting the position of the
transient to host redshift and then back to all possible the image
positions), pre-/re-covery of images predicted to lead and lag
the alert transient, corroborating (by detecting
at least one additional image) or rebutting by setting limits,
phasing of the light curves of the corroborated images, 
verifying spectroscopically that the transient is indeed
a supernova, and reconstruction of the lens model adding
data for the SN and its host obtained following the alert.
Many of these require ancillary observations to be carried out
with other telescopes.  If SN 2025wny is any indication,
there will be ample opportunity to do so.

\begin{acknowledgments}
  
  The authors thank Gautham Narayan and Mario Juri\'c for helpful
  suggestions.  They thank Michael Wood-Vasey for encouragement.  MZL
  thanks the MIT Undergraduate Research Opportunities Program for
  support.

\end{acknowledgments}

\software{isit4or2or1v1.0~ doi:105281/zenodo.20086659}

\appendix

\section{Timeline:  Observations of SN 2025wny and Associated Postings}\label{app:timeline}

Table \ref{tab:timeline} gives the timeline of  observations and postings that contributed
to the confirmation of SN 2025wny as a quadruply lensed quasar and that
will ultimately contribute to better models than the one presented here.

The  host galaxy PS 1J0716+3821 had already previously identified as quadruply
lensed \citep{Canameras}, and the redshift of the lens galaxy had been reported
by the DESI survey \citep{Bellm} to be to be $z_{G1} = 0.3754$.
\vfill
\eject

\begin{deluxetable*}     {lll}\label{tab:timeline}
\tablecolumns{3}
\tablewidth{920pt}
\tablecaption{Timeline:  Observations of SN2025wny and Associated Postings}
  \tablehead{
    \colhead{date} &
    \colhead{Facilty} &    
    \colhead{Event} \\
  }
  \startdata
Aug 18$\rightarrow$ & ZTF$^a$  & field included in survey schedule \\
Aug 27 &     ZTF &              pre-covery exposure taken \\
Aug 29 &    ZTF  &              transient identified ZTF25abnjznp \\
Sep 01 &    TNS$^b$  &       reported as  GOTO25gtq  \\
Sep 02$\rightarrow$ &    LT$^c$  &  direct imaging campaign begun \\
Sep 20  &    LT    &               images B \& C detected \\
Sep 26  &    NOT$^d$ &               spectrum shows SN-like features \\
Sep 29$\rightarrow$ &  Lulin$^e$ &           direct campaign begun \\
Sep 29  &    Lulin      &          images A-D detected \\
Oct 03  &     TNS      &          designated AT 2025wyn ; images A-D reported \\
Oct 15$\rightarrow$  &    COLIBRI$^g$ &   direct campaign begun \\
Oct 15$\rightarrow$  &    Maidanak$^h$ &   direct campaign begun \\
Oct 22 &   HST$^i$     & program 17611 epoch 1 \\
Oct 22 &   JWST$^j$    & program 5564 triggered \\
Oct 27   &     arXiv$^k$   &     Taubenberg et al.\ 2510.21694 posted \\
Oct 27   &    arXiv$^l$     &    Johansson et al.\ 2510.23533 posted  \\
\enddata
\tablecomments{
  $^a$Zwicky Transient Facility (https://www.ztf.caltech.edu);
  $^b$Transient Name Server (https://www.wis-tns.org);
  $^c$Liverpool 2.2-m Telescope (https://telescope.livjm.ac.uk);
  $^d$Nordic Optical Telescope with ALFOSC spectrometer (https://www.not.iac.es);
  $^e$Lulin Observatory 1-m Telescope (https://tara.tw/lot-2mt.html)
  $^g$COLIBRI Telescope \citet{BasaII} ;
  $^h$Maidanak Observatory 1.5-m telescope (http://www.maidanak.uz);
  $^i$Hubble program 17611 (https://www.stsci.edu/hst-program-info);
  $^j$JWST program 5564 (https://www.stsci.edu/jwst/science-execution);  
  $^k$\citet{Taubenberger};
  $^l$J. Johannson et al. (2025).
  }
\end{deluxetable*}



\section{Interpretation of the LBT SN exposures with the CFHT Witt-Wynne model}\label{app:WittWynne}

Figure \ref{fig:hostSNpred} shows our Witt-Wynne geometric lens model
overlayed on the CFHT figure on which the host positions were
measured.  The parameters for the associated singular isothermal elliptical potential
are:
{\parskip=0pt
\begin{itemize}
\item ellipticity of potential -- $\epsilon = 0.29$;
\item semi-major axis of potential -- $a = 1\farcs77$.
\item orientation of long axis of potential --  P.A. = $22^\circ$ East of North.
\end{itemize}
}
The long axis of Wynne's ellipse is perpendicular to the long axis of the potential, but has the same semi-major axis.

Table \ref{tab:modSTUFF} gives predictions for the host positions relative
to the center of the model, and for the time delays and magnifications
at those positions.  The measured positions for lensing galaxies $G1$
and $G2$, again with respect to the center of the model, are also
given.  A flat $\Lambda$CDM cosmology was used with $H_0 = 70$ km/s
and $\Omega_m$ = 0.3.

We saw in Figure \ref{fig:hostSNpred} that the Witt-Wynne model did a
reasonably good job of propagating each of the four SN images back
to the host redshift and then forward to the remaining images.
According to that model,  SN image A is the trailing image, 20, 30 and 175
days behind images C, B and D, respectively.\footnote{Note the small differences from the values predicted in Table \ref{tab:modSTUFF} using the host
image positions} ``Pre-covery'' of the
brightening seen in A would be challenging given the Right Ascension
of SN 2025wny.  The best opportunity for ``pre-covery'' might be for
image D, since it would still have been observable in March 2025.

Almost as disappointing as the long delay is the large ellipticity for
the potential.  Using equation (E8) of \citet{LuhtaruI}, the most
closely matching elliptical mass distribution would have an
ellipticity of $\sim 0.36$, exceedingly flat and unlike
any known quad. The same authors also show that the same image
configuration would be produced by a isothermal sphere with external
shear $\gamma = 0.14$ equal to half the ellipticity of the potential.
As Luhtaru et al. show, this is high even for know quadruply lensed
quasars.

If we attribute all of the elongation of the potential to shear,
instead of lens ellipticity, the delays are shorter by a factor of
two, dashing the hope of ``pre-covering'' the rise of image D.

The magnifications for our model are quite low.
If the true magnifications are substantially higher, one might
expect to see elongation of the host images with HST and JWST once 
the supernova has faded.  We anticipate the resolution of the
questions posed in this appendix when the results of the ongoing observing
campaigns described in the discovery papers are presented.






\vfill
\eject
\bibliography{schechter}{}

\begin{thebibliography}{}
\expandafter\ifx\csname natexlab\endcsname\relax\def\natexlab#1{#1}\fi
\providecommand{\url}[1]{\href{#1}{#1}}
\providecommand{\dodoi}[1]{doi:~\href{http://doi.org/#1}{\nolinkurl{#1}}}
\providecommand{\doeprint}[1]{\href{http://ascl.net/#1}{\nolinkurl{http://ascl.net/#1}}}
\providecommand{\doarXiv}[1]{\href{https://arxiv.org/abs/#1}{\nolinkurl{https://arxiv.org/abs/#1}}}

\bibitem[{S. {Basa} {et~al.}(2026){Basa}, {Lee}, {Watson}, {Dolon}, {Floriot},
  {Atteia}, {Dornic}, {Lugo-Ibarra}, {Figueroa}, {Langarica}, {Valentin},
  {Ageron}, {Agneray}, {Alvarez Nunez}, {Angulo-Valdez}, {Antier}, {Auphan},
  {Baumann}, {Bautista}, {Becerra}, {Benahmed}, {Benamar}, {Blanpain},
  {Boulade}, {Bounab}, {Boy}, {Butler}, {Cadena Zepeda}, {Cuevas}, {de Ugarte
  Postigo}, {Delisle}, {Devigny}, {Ducoin}, {Fortin}, {Fuentes-Fernandez},
  {Gaiti}, {Garcia-Garcia}, {Gallais}, {Gill}, {Globus}, {Guisa}, {Kajfasz},
  {Lafforgue}, {Langlois}, {Larrieu}, {Landa}, {Lecubin}, {Lopez-Camara},
  {Lopez Angeles}, {Lombardo}, {Magnani}, {Mandarakas}, {Malgoyre}, {Mathon},
  {Moreno Mendez}, {Moreau}, {Nouvel de la Fleche}, {Ochoa}, {Ortiz},
  {Pedrayes-Lopez}, {Pereyra}, {Provost}, {Ramon}, {Rakotondrainibe},
  {Ronayette}, {Ruiz Diaz-Soto}, {Sanchez Alvarez}, {Schneider}, {Secroun},
  {Striebieg}, {Tinoco}, {Tourner-Sylvain}, {Valenzuela}, \&
  {Vincent}}]{BasaII}
{Basa}, S., {Lee}, W.~H., {Watson}, A.~M., {et~al.} 2026,
  \bibinfo{title}{{COLIBRI (SVOM/FM-GFT): Instrumentation and Performances on
  the SVOM Alerts},} arXiv e-prints, arXiv:2604.24259.
\newblock \doarXiv{2604.24259}

\bibitem[{E.~C. {Bellm} {et~al.}(2019){Bellm}, {Kulkarni}, {Graham}, {Dekany},
  {Smith}, {Riddle}, {Masci}, {Helou}, {Prince}, {Adams}, {Barbarino},
  {Barlow}, {Bauer}, {Beck}, {Belicki}, {Biswas}, {Blagorodnova}, {Bodewits},
  {Bolin}, {Brinnel}, {Brooke}, {Bue}, {Bulla}, {Burruss}, {Cenko}, {Chang},
  {Connolly}, {Coughlin}, {Cromer}, {Cunningham}, {De}, {Delacroix}, {Desai},
  {Duev}, {Eadie}, {Farnham}, {Feeney}, {Feindt}, {Flynn}, {Franckowiak},
  {Frederick}, {Fremling}, {Gal-Yam}, {Gezari}, {Giomi}, {Goldstein},
  {Golkhou}, {Goobar}, {Groom}, {Hacopians}, {Hale}, {Henning}, {Ho}, {Hover},
  {Howell}, {Hung}, {Huppenkothen}, {Imel}, {Ip}, {Ivezi{\'c}}, {Jackson},
  {Jones}, {Juric}, {Kasliwal}, {Kaspi}, {Kaye}, {Kelley}, {Kowalski},
  {Kramer}, {Kupfer}, {Landry}, {Laher}, {Lee}, {Lin}, {Lin}, {Lunnan},
  {Giomi}, {Mahabal}, {Mao}, {Miller}, {Monkewitz}, {Murphy}, {Ngeow},
  {Nordin}, {Nugent}, {Ofek}, {Patterson}, {Penprase}, {Porter}, {Rauch},
  {Rebbapragada}, {Reiley}, {Rigault}, {Rodriguez}, {van Roestel}, {Rusholme},
  {van Santen}, {Schulze}, {Shupe}, {Singer}, {Soumagnac}, {Stein}, {Surace},
  {Sollerman}, {Szkody}, {Taddia}, {Terek}, {Van Sistine}, {van Velzen},
  {Vestrand}, {Walters}, {Ward}, {Ye}, {Yu}, {Yan}, \& {Zolkower}}]{Bellm}
{Bellm}, E.~C., {Kulkarni}, S.~R., {Graham}, M.~J., {et~al.} 2019,
  \bibinfo{title}{{The Zwicky Transient Facility: System Overview, Performance,
  and First Results},} \pasp, 131, 018002, \dodoi{10.1088/1538-3873/aaecbe}

\bibitem[{R. {Blandford} \& R. {Narayan}(1986){Blandford} \&
  {Narayan}}]{Blandford}
{Blandford}, R., \& {Narayan}, R. 1986, \bibinfo{title}{{Fermat's Principle,
  Caustics, and the Classification of Gravitational Lens Images},} \apj, 310,
  568, \dodoi{10.1086/164709}

\bibitem[{R. {Ca{\~n}ameras} {et~al.}(2020){Ca{\~n}ameras}, {Schuldt}, {Suyu},
  {Taubenberger}, {Meinhardt}, {Leal-Taix{\'e}}, {Lemon}, {Rojas}, \&
  {Savary}}]{Canameras}
{Ca{\~n}ameras}, R., {Schuldt}, S., {Suyu}, S.~H., {et~al.} 2020,
  \bibinfo{title}{{HOLISMOKES. II. Identifying galaxy-scale strong
  gravitational lenses in Pan-STARRS using convolutional neural networks},}
  \aap, 644, A163, \dodoi{10.1051/0004-6361/202038219}

\bibitem[{L.~R. {Ecker} {et~al.}(2026){Ecker}, {Schweinfurth}, {Saglia},
  {Deng}, {Suyu}, {Saulder}, {Snigula}, {Bender}, {Ca{\~n}ameras}, {Chen},
  {Galan}, {Halkola}, {Mamuzic}, {Melo}, {Schuldt}, \& {Taubenberger}}]{Ecker}
{Ecker}, L.~R., {Schweinfurth}, A.~G., {Saglia}, R., {et~al.} 2026,
  \bibinfo{title}{{HOLISMOKES XX. Lens models of binary lens galaxies with five
  images of Supernova Winny},} arXiv e-prints, arXiv:2602.16620,
  \dodoi{10.48550/arXiv.2602.16620}

\bibitem[{C. {Falor} \& P.~L. {Schechter}(2022){Falor} \& {Schechter}}]{Falor}
{Falor}, C., \& {Schechter}, P.~L. 2022, \bibinfo{title}{{The Quadruple Image
  Configurations of Asymptotically Circular Gravitational Lenses},} A.J., 164,
  120, \dodoi{10.3847/1538-3881/ac80bc}

\bibitem[{J. {Johansson} {et~al.}(2025){Johansson}, {Perley}, {Goobar}, {Wise},
  {Qin}, {McGrath}, {Schulze}, {Lemon}, {Gangopadhyay}, {Tsalapatas},
  {Andreoni}, {Bellm}, {Bloom}, {Dekany}, {Dhawan}, {Fremling}, {Graham},
  {Groom}, {Gruen}, {Hall}, {Kasliwal}, {Laher}, {Lunnan}, {Mahabal}, {Miller},
  {M{\"o}rtsell}, {Nordin}, {Osman Hjortlund}, {Rich}, {Riddle}, {Singh},
  {Sollerman}, {Townsend}, \& {Yan}}]{Johansson}
{Johansson}, J., {Perley}, D.~A., {Goobar}, A., {et~al.} 2025,
  \bibinfo{title}{{Discovery of SN 2025wny: a Strongly Gravitationally Lensed
  Superluminous Supernova at z = 2.01},} arXiv e-prints, arXiv:2510.23533,
  \dodoi{10.48550/arXiv.2510.23533}

\bibitem[{T.~L. {Killestein}(2026){Killestein}}]{Killestein}
{Killestein}, T.~L. 2026, \bibinfo{title}{{StarHash: unique, memorable, and
  deterministic names for astronomical objects},} arXiv e-prints,
  arXiv:2603.29584, \dodoi{10.48550/arXiv.2603.29584}

\bibitem[{C. {Lemon} {et~al.}(2026){Lemon}, {Goobar}, {Johansson},
  {M{\"o}rtsell}, {Schulze}, {Andreoni}, {Bochenek}, {Brennan}, {Busmann},
  {Coughlin}, {Das}, {Dhawan}, {Fremling}, {Gangopadhyay}, {Gruen}, {Hall},
  {Ho}, {Kasliwal}, {Perley}, {Rigault}, {Schroeder}, {Smith}, {Sollerman},
  {Somalwar}, {Stein}, {Thorp}, {Townsend}, {Wise}, {Yan}, {Arendse}, {Bellm},
  {Chen}, {Drake}, {Masci}, {Purdum}, {Smith}, {Hinkle}, {Rivera-Thorsen},
  {Shappee}, {Tucker}, {Aguilar}, {Ahlen}, {Aldering}, {Benzvi}, {Bianchi},
  {Brooks}, {Claybaugh}, {de la Macorra}, {Della Costa}, {Dey}, {Doel},
  {Flaugher}, {Font-Ribera}, {Forero-Romero}, {Gazta{\~n}aga}, {Gontcho},
  {Gutierrez}, {Huterer}, {Ishak}, {Jimenez}, {Joyce}, {Juneau}, {Kehoe},
  {Kim}, {Kirkby}, {Kisner}, {Kremin}, {Lahav}, {Landriau}, {Le Guillou},
  {Levi}, {Manera}, {Meisner}, {Miquel}, {Moustakas}, {Nadathur}, {O'Connor},
  {Palanque-Delabrouille}, {Palmese}, {Percival}, {P{\'e}rez-R{\`a}fols},
  {Poppett}, {Prada}, {Rossi}, {Sanchez}, {Schlegel}, {Schubnell}, {Shafieloo},
  {Silber}, {Sprayberry}, {Tarl{\'e}}, {Weaver}, \& {Zou}}]{lemon}
{Lemon}, C., {Goobar}, A., {Johansson}, J., {et~al.} 2026, \bibinfo{title}{{A
  Natural $rsim 100\times$ Telescope: Discovery of the Strongly Lensed Type II
  SN 2025mkn at $z=1.37$},} arXiv e-prints, arXiv:2604.07983,
  \dodoi{10.48550/arXiv.2604.07983}

\bibitem[{R. {Luhtaru} {et~al.}(2021){Luhtaru}, {Schechter}, \& {de
  Soto}}]{LuhtaruI}
{Luhtaru}, R., {Schechter}, P.~L., \& {de Soto}, K.~M. 2021,
  \bibinfo{title}{{What Makes Quadruply Lensed Quasars Quadruple?},} \apj, 915,
  4, \dodoi{10.3847/1538-4357/abfda1}

\bibitem[{P.~L. {Schechter} \& R. {Luhtaru}(2024){Schechter} \&
  {Luhtaru}}]{LuhtaruII}
{Schechter}, P.~L., \& {Luhtaru}, R. 2024, \bibinfo{title}{{Witt's Hyperbola Is
  Both Predicted and Observed to Pass Close to the Lensing Galaxies in
  Quadruple Quasars},} \apj, 975, 62, \dodoi{10.3847/1538-4357/ad7f4d}

\bibitem[{C.~J. {Storfer} {et~al.}(2026){Storfer}, {Wong}, {Acebron}, {Grillo},
  {Hoogendam}, {Huang}, {Jones}, {Magnier}, {Mandel}, {Ratier-Werbin}, {Rubin},
  {Shappee}, \& {Soler-Perez}}]{Storfer}
{Storfer}, C.~J., {Wong}, K.~C., {Acebron}, A., {et~al.} 2026,
  \bibinfo{title}{{Supernova 2025wny: High-angular resolution Keck/NIRC2
  observations and preliminary lens modeling},} arXiv e-prints,
  arXiv:2604.02418, \dodoi{10.48550/arXiv.2604.02418}

\bibitem[{S. {Taubenberger} {et~al.}(2025){Taubenberger}, {Acebron},
  {Ca{\~n}ameras}, {Chen}, {Galan}, {Grillo}, {Melo}, {Schuldt},
  {Schweinfurth}, {Suyu}, {Aldering}, {Aryan}, {Lee}, {Mamuzic}, {Millon},
  {Reynolds}, {Sergeyev}, {Asfandiyarov}, {Basa}, {Blondin}, {Burkhonov},
  {Christensen}, {Courbin}, {Ehgamberdiev}, {Killestein}, {Mattila},
  {Shaymanov}, {Shu}, {Xu}, {Yang}, {Gruen}, {Pierel}, {Storfer}, {Tran},
  {Wong}, {Becerra}, {Dornic}, {Ducoin}, {Globus}, {Guti{\'e}rrez}, {Jiang},
  {Kuncarayakti}, {L{\'o}pez-C{\'a}mara}, {Lundqvist}, {Magnani}, {Moreno
  M{\'e}ndez}, {Schneider}, \& {Vogl}}]{Taubenberger}
{Taubenberger}, S., {Acebron}, A., {Ca{\~n}ameras}, R., {et~al.} 2025,
  \bibinfo{title}{{HOLISMOKES XIX: SN 2025wny at $z=2$, the first strongly
  lensed superluminous supernova},} arXiv e-prints, arXiv:2510.21694,
  \dodoi{10.48550/arXiv.2510.21694}

\bibitem[{C. {Vanderriest} {et~al.}(1986){Vanderriest}, {Wlerick}, {Lelievre},
  {Schneider}, {Sol}, {Horville}, {Renard}, \& {Servan}}]{Vanderriest}
{Vanderriest}, C., {Wlerick}, G., {Lelievre}, G., {et~al.} 1986,
  \bibinfo{title}{{La variabilite du mirage gravitationnel P.G. 1115+080.},}
  \aap, 158, L5

\bibitem[{L. {Weisenbach} {et~al.}(2021){Weisenbach}, {Schechter}, \&
  {Pontula}}]{Weisenbach}
{Weisenbach}, L., {Schechter}, P.~L., \& {Pontula}, S. 2021,
  \bibinfo{title}{{``Worst-case'' Microlensing in the Identification and
  Modeling of Lensed Quasars},} \apj, 922, 70, \dodoi{10.3847/1538-4357/ac2228}

\bibitem[{R.~J. {Weymann} {et~al.}(1980){Weymann}, {Latham}, {Angel}, {Green},
  {Liebert}, {Turnshek}, {Turnshek}, \& {Tyson}}]{Weymann}
{Weymann}, R.~J., {Latham}, D., {Angel}, J. R.~P., {et~al.} 1980,
  \bibinfo{title}{{The triple QSO PG1115 + 08: another probable gravitational
  lens},} \nat, 285, 641, \dodoi{10.1038/285641a0}

\bibitem[{J. {Wise} {et~al.}(2025){Wise}, {Perley}, {Goobar}, {Johansson}, \&
  {McGrath}}]{Wise}
{Wise}, J., {Perley}, D., {Goobar}, A., {Johansson}, J., \& {McGrath}, Z. 2025,
  \bibinfo{title}{{Liverpool Telescope Discovery of Multiple Images from
  AT2025wny (ZTF25abnjznp/GOTO25gtq)},} Transient Name Server AstroNote, 296, 1

\bibitem[{H.~J. {Witt}(1996){Witt}}]{Witt}
{Witt}, H.~J. 1996, \bibinfo{title}{{Using Quadruple Lenses to Probe the
  Structure of the Lensing Galaxy},} \apjl, 472, L1, \dodoi{10.1086/310358}

\bibitem[{R.~A. Wynne \& P.~L. Schechter(2018)Wynne \& Schechter}]{Wynne}
Wynne, R.~A., \& Schechter, P.~L. 2018, Robust modeling of quadruply lensed
  quasars (and random quartets) using Witt's hyperbola, \doarXiv{1808.06151}

\end{thebibliography}
\bibliographystyle{aasjournalv7}



\end{document}